# Defects Evolution and Mg Segregation in Mg-implanted GaN with Ultra-High-Pressure Annealing

Y. Wang[1], K. Huynh[1], M.E. Liao[1], J. Tweedie[2], P. Reddy[2], M.H. Breckenridge[3], R. Collazo[3], Z. Sitar[3], M. Bockowski[4], X. Huang[5], M. Wojcik[5], M.S. Goorsky[1]

1. Department of Materials Science and Engineering, University of California, Los Angeles, Los Angeles, California 90095, USA

2. Adroit Materials, Inc, Apex, NC 27518, USA

3. Department of Material Science and Engineering, North Carolina State University, Raleigh, NC 27606, USA

4. Institute of High Pressure Physics Polish Academy of Sciences, Warsaw 01-142, Poland

5. Advanced Photon Source, Argonne National Lab, Lemont, IL 60439, USA

**Abstract**

Annealing Mg-implanted homoepitaxial GaN at temperatures at or above 1400 °C eliminates the formation of inversion domains and leads to improved dopant activation efficiency. Extended defects in the form of inversion domains contain electrically inactive Mg after post-implantation annealing at temperatures as high as 1300 °C (one GPa $N_2$ overpressure), which results in a low dopant activation efficiency. Triple axis X-ray data show that the implant-induced strain is fully relieved after annealing at 1300 °C for 10 min, indicating that the strain-inducing point defects formed during implantation have reconfigured. However, annealing at temperatures of 1400 °C to 1500 °C (also one GPa $N_2$ overpressure) eliminates the presence of the inversion domains. Annealing at these higher temperatures and for longer time does not have any further impact on the strain state. While residual defects, such as dislocation loops, still exist after annealing at and above 1400 °C, chemical analysis at the dislocation loops shows no sign of Mg segregation. Meanwhile, an overall decreasing trend in the dislocation loop size and density is observed after annealing at the higher temperatures and longer times. Earlier work [1] addressing electrical measurements of these types of samples showed that annealing at 1400 °C leads to a dopant activation efficiency that is an order of magnitude higher than that observed at 1300 °C. This work complements the earlier work by identifying the microscopic defects (inversion domains) which incorporate Mg, and points to the benefits, in terms of defect density and p-type dopant activation, of using higher temperatures (≥ 1400 °C) annealing cycles to activate Mg in GaN.

**Introduction**

Vertical GaN power devices have emerged as promising candidates for next-generation high power applications due to superior material properties such as high breakdown voltage, low on-resistance, and high mobility compared to devices based on Si and SiC [2, 3]. GaN-based power switches offer higher voltage power with significantly higher efficiencies at a much smaller form factor. A technological limitation of GaN based vertical devices (as well as other GaN devices that require high p-type doping concentrations) has been the inability to achieve planar high p-type doping. Ion implantation has been very effective in selective area p-type doping for Si- and SiC-based technology [4, 5]. However, p-type selective area doping with ion implantation in GaN remains a challenging task.

The activation of the implanted p-type dopants requires high temperature annealing. The targeted annealing temperature for GaN, based on the experience from other semiconductors, is ~ 2/3 the melting point, corresponding to a temperature of ~1500 °C. However, GaN is thermodynamically unstable at T > 850 °C under atmospheric pressure. Various efforts have aimed to avoid GaN decomposition during high temperature processing, such as using dielectric capping materials (e.g., AlN) with a rapid heating source [6, 7, 8] or applying high $N_2$ pressure without using any caps [9, 10, 11]. After annealing, residual defects, such as dislocation loops, inversion domains, and stacking faults have been observed in the implanted region [12, 13, 14]. It is believed that these defects form from the point defects that are introduced through implantation and that different defects are stable at different temperatures. For example, we have shown earlier that stacking faults exist after short time (10 minutes) annealing at 1300 °C but are not present after extended annealing (100 minutes). Dislocation loops are stable at the same temperature [15]. After implantation and annealing at 1300 °C (10 minutes), cubic domains (~ 10 nm regions embedded in the wurtzite lattice) are also present [15], which indicates a significant amount of lattice reconfiguration/rearrangement occurs at 1300 °C and likely at higher temperatures as well.

A high quality homoepitaxial structure provides important benefits to study the evolution of the extended defects. Due to the lack of inexpensive high quality, and large size native substrate, GaN epitaxial layers for prior p-type ion implantation studies were typically grown on highly mismatched sapphire substrates. Our earlier work showed a large number of defects, partially due to the high intrinsic defect density (> $10^8$ cm$^{-2}$) in such heteroepitaxial structures, remained after the activation anneal, which limits subsequent device performance [16]. The recent developments in bulk growth techniques such as hydride vapor phase epitaxy and ammonothermal growth have led to bulk GaN substrates with substantially lower defect density (< $10^6$ cm$^{-2}$) [17, 18]. Using homoepitaxial GaN layers grown on ammonothermal substrates, we have developed X-ray topography techniques and combined these with electron microscopy and high resolution X-ray diffraction to understand how implant-induced strain and defects evolve during subsequent high temperature processing steps [15]. The next step, discussed here, is to understand how these extended defects affect dopant activation. For example, in the current literature, Mg segregation at inversion domains has been observed in highly Mg-doped epitaxial layers with Mg concentration above the $10^{19}$ cm$^{-3}$ level [19, 20, 21]. It was found that inversion domains play a role in the free carrier reduction in highly Mg-doped GaN. Recently, Kumar et al., showed Mg-rich clusters started to form in Mg implanted GaN after annealing at 1100 °C and remained after annealing at ~1350 °C [22, 23]. Iwata et al., further discovered Mg segregation at the boundary of the inversion domains in Mg implanted GaN after annealing at ~1300 °C [14]. Interestingly, the two implant studies both showed evidence of Mg segregation in extended defects after annealing at approximately the same temperature. However, the question remains how these inversion domains affect the dopant activation efficiency and if they remain stable after annealing at higher temperatures and/or for longer times. In our earlier study, post-implantation annealing at 1300 °C for 10 min and 100 min results in changes in both defect concentration and defect structures [15]. Therefore, when the annealing temperature approaches ~2/3 the GaN melting point, it is expected that there would be further lattice rearrangement and reconfiguration of the defect structures.

Based on the defect characterization method with X-ray scattering and electron microscopy techniques that we developed previously [24, 25], this study aims to assess (i) the key residual defects after annealing that have a significant impact on the dopant activation efficiency and (ii) how to remove them. These results are anticipated to guide the key processing steps and requirements to achieve high activation efficiency p-type doping for vertical GaN devices.

**Experimental**

GaN epitaxy films with a thickness of 2.5 μm were grown via metal organic chemical vapor deposition (MOCVD) on high quality ammonothermal GaN substrates [1, 15]. The magnesium ions with a box profile were implanted at room temperature using six different accelerating voltages (350 keV, 200 keV, 150 keV, 100 keV, 60 keV, and 25 keV) at a total dose level of $1\times10^{15}$ cm$^{-2}$. Using the SRIM simulation software [26], the maximum concentration was calculated to be $1.8\times10^{19}$ cm$^{-3}$ from the surface to a depth of ~500 nm. Post-implantation annealing was performed on the implanted wafers at temperatures of 1300 °C, 1400 °C, and 1500 °C for a duration of either 10 minutes or 100 minutes and so will be referred to as, for example, the 1300°C; 10 min., sample. An ultra-high $N_2$ pressure of one GPa was applied during the annealing sequence to prevent GaN from decomposition without using a cap [1, 15].

Structural characterization was performed by a combination of X-ray scattering and electron scattering techniques. The lattice distortions and crystalline quality were assessed by Triple-axis X-ray diffraction (TAXRD) and X-ray topography (XRT). TAXRD measurements were performed using a Jordan Valley (Bruker) D1 diffractometer, with an incident beam mirror to produce a parallel beam, followed by a Si (220) channel cut collimator (Cu K$\alpha_1$ radiation). The scattered beam optics included a Si (220) channel cut crystal [15, 16, 27]. Synchrotron double crystal X-ray topography measurements were performed at the 1-BM Beamline of the Advanced Photon Source, Argonne National Laboratory with a photon energy of 8.05 keV. The first crystal was a highly asymmetric Si (333) beam expander and the sample was oriented for diffraction of the (11$\bar{2}$4) reflection in the glancing incidence geometry [24, 25]. Topography images were taken by exposing at different positions on the rocking curve and recorded separately on different films (hereinafter referred to as single exposure images). Information of the specific post-annealing defects was obtained using Transmission Electron Microscope (TEM). TEM samples roughly 100 nm thick were made using an FEI Nova 600 Dual Beam Focused Ion Beam (FIB) and further thinned by a gentle Ar+ ion beam with 0.3 kV incident energy to remove any FIB induced damage layer. Scanning Transmission Electron Microscope (STEM) images using bright field (BF) or high-angle annular dark field (HAADF) detectors, and High-Resolution TEM images were taken using the FEI Talos and Argonne Chromatic and Aberration-Corrected (ACAT) TEM respectively at the Center for Nanoscale Materials, Argonne National Laboratory [28].

**Results and discussions**

The triple axis X-ray ω:2θ line scans near GaN (0004) peak of all the samples are shown in Figure 1a. The presence of the intensity with fringes to the left of the main peak in the as-implanted sample is due to lattice distortion as a result of the implant-induced point defect formation [29, 30, 31]. After annealing, in all cases, the implantation-induced strain is completely relieved. Previously, we showed that annealing under a similar condition at 1300 °C for 10 min removed the implantation-induced strain for a single implant (Mg: 100 keV) at a dose level of $2\times10^{14}$ cm$^{-2}$ [15]. In this work, the dose level is five times of that previous work, and the same annealing condition is still sufficient to relieve the strain. Meanwhile, annealing at higher temperatures and longer time does not have any further impact on the strain state. Thus in all cases, the point defects, which were responsible for the strain have reformed into more stable configurations that can include extended defects. The residual defect structures after annealing were first screened using X-ray topography. Figure 1(b-e) show the single exposure curves X-ray topography images, taken at the peak of the (11$\underline{2}$4) rocking curves, from the as-implanted sample and samples annealed at 1300 °C, 1400 °C, and 1500 °C. In all cases, dot-shape defects are observed, which correspond to individual threading dislocations. The dislocation density is on the order of $10^4$ cm$^{-2}$, which is consistent with the dislocation density level in ammonothermal GaN reported in the current literature [18]. The as-implanted and the 1500 °C annealed samples show similar characteristics: only individual threading dislocations with no extended defects are present. The 1400 °C sample consists of isolated loop-shaped defects that do not diffract at the same angle as the rest of the material. Separate images capture the loops diffracting at an angular difference of ~8 arcsec away from the bulk of the GaN, quantifying the small but measurable local lattice distortion at the loop defects. The amount of the local lattice distortion here is comparable to an earlier work from our group, using laboratory XRT equipment, to determine the localized tilt (a few arcsecs) around micropipe defects in SiC substrates [32]. The 1300 °C sample, on the other hand, exhibits large non-diffracting features that are not observed in the other samples; (arrow in figure 1c). Details of the extended defects were further characterized using TEM/STEM images. Figure 2 shows the STEM two-beam condition bright field images taken under different diffraction conditions. Figure 2 a-d are taken under $\boldsymbol{g} = <0002>$ to show, for example, defects with a screw component and figure 2 e-h are taken under $\boldsymbol{g} = <11\bar{2}0>$ to show defects with an edge component. The four samples can be divided into two groups based on the extended defects observed. The samples annealed at

1400 °C and 1500 °C showed very similar defect structures. The prominent residual defects after annealing at and above 1400 °C are dislocation loops. Under $g$ = <0002>, the defects observed possess faint contrast from the dislocation loops. The faint contrast could come from the part of the loop having a screw component. Under $g$ = <11$\bar{2}$0>, the loop defects have stronger contrast as observed in the three samples annealed at 1400 °C and 1500 °C, suggesting that the loops produced at these elevated temperatures are dominated by edge characteristics. Most of the loops are not completely visible. Because those loops lie on crystallographic planes that are not perfectly parallel to the sample orientation, i.e. {10$\underline{1}$0}, only a portion of the loop is oriented to show contrast, an example is given with a blue arrow in figure 2f. In the extreme case, when the loop is oriented perpendicular to the sample orientation, for example {11$\underline{2}$0}, it will show up as a bright line surrounded by sharp contrast on both sides, an example is given with a red arrow in figure 2f. The contrast is caused by lattice distortion near the dislocation loop.

The 1300 °C 10 min sample exhibits an additional, different defect structure than the others. Triangular and trapezoidal shape defects were observed under $g$ = <0002>, these defects did not exist in the three samples annealed at higher temperatures. A magnified STEM bright field image taken under $g$ = <0002> for the 1300 °C 10 min sample is shown in figure 3a. Examples of the triangular and trapezoidal shape defects are circled in red and highlighted in the blue box. The triangular defects showed structural characteristics that those of pyramidal inversion domains (PIDs). The base of the PIDs is along the {0001} plane and the sidewalls (highlighted with dotted lines) are along the {11$\underline{2}$3} planes, inclined at ~47°, as shown by high-resolution TEM image in figure 3b. The lattice is distorted inside the PID when compared to the surrounding GaN matrix. An additional atomic layer (highlighted with a green box) was observed near the base of the pyramid (basal plane of GaN), causing lattice bending of the following ten layers. For an individual PID (the STEM HAADF image is shown in figure 3c), the Energy Dispersive X-Ray Analysis (EDX) map in figure 3d shows the Mg signal at the PID. A line profile is generated by integrating the intensity in the yellow box in figure 3d and shown in figure 3e. From the EDX map and the line profile, an increased Mg signal is observed at the base of the PID, where the extra layer of atoms is located, with a decrease in the Ga signal at the same position. This indicates that there is possibly a compound formation including the Mg and N in the PIDs. For the trapezoidal shape defects, we also observed facets along the {11$\underline{2}$3} planes on the edge, shown in figure 4a(highlighted with orange lines). The EDX map and line profile from the trapezoidal shape defects also show increased Mg signal and reduced Ga signal along the basal plane, shown in Figures 4b and 4c, which is similar to what was observed in the PIDs. Therefore, we consider these trapezoidal shape defects as trapezoidal inversion domains (TIDs). In fact, the center part of the largest TID consists of a series of small aligned pyramids, as shown in figure 3a. We speculate that the TIDs are the initial stages of development and will eventually evolve into the PIDs. Thus, it appears that both the PIDs and TIDs defects contain electrically inactive Mg atoms and thus are the cause of the reduction in Mg activation efficiency. In the current literature, PIDs with signs of Mg segregation have been observed in both highly Mg-doped epitaxial layers with Mg concentration above the $10^{19}$ cm$^{-3}$ level [19-21] as well as in Mg implanted GaN after annealing at 1350 °C [23]. In MOVPE epitaxial Mg-doped p-GaN, Vennegues et al., hypothesized that introducing a higher level of Mg in GaN results in the formation of $Mg_3N_2$ where inactive Mg atoms do not substitute Ga sites but are accommodated into other sites [20]. Our data supports this hypothesis and shows that a similar interaction between Mg and N occurred in Mg implanted GaN after annealing at 1300 °C. The base of PIDs can accommodate the Mg and prevent it from becoming electrically active. However, in those epitaxially grown structures, it is difficult to remove the PIDs once they are formed. To suppress the formation of the PIDs, a low-temperature growth method is needed, such as metal-modulated epitaxy [33]. Unlike the case for annealing the ion implantation damage, epitaxial growth of highly Mg-doepd GaN at higher temperatures ($\geq$ 1400 °C) is problematic because ammonia (the source) dissociates and forms a high concentration of hydrogen, which passivates Mg and form electrically inactive Mg-H complexes [34]. In the case of Mg implanted GaN, the current literature have not shown how to remove or avoid the PIDs. However, the PIDs and TIDs are not observed in any of the samples annealed at and above 1400 °C in this work. This suggests that the PIDs and TIDs either do not form or are dissolved at high temperatures ($\geq$ 1400 °C) and are formed only at temperatures ~ 1300 °C as presented here and also described in works by Kumar et al., [22, 23] for implantation, and in works by Vennegues et al., [19, 20] for epitaxial structures in which high concentrations of Mg are present. This provides insight for improving dopant activation efficiency by annealing at temperatures above 1400 °C. Recently, Breckenridge et al., [1] reported the electrical data from a sister set of the samples annealed up to 1400 °C under similar conditions. Their results showed that annealing at 1400 °C leads to higher dopant activation (~10% at room temperature, ~100% at 500 °C, because the Mg level is not that shallow – at ~ $E_v$+ 0.2 eV), compared to annealing at 1300 °C (<1% at room temperature, ~40% at 500 °C). Our study complements the earlier study from Breckenridge et al., by providing the structural analysis of the residual defects and showing the Mg segregation in PIDs and TIDs are detrimental to dopant activation and can be avoided if the annealing temperature is ≥ 1400 °C.

As shown before, the prominent residual defects after annealing at and above 1400 °C are dislocation loops. High resolution TEM image from one of the small loops in the 1400 °C 10 min sample is shown in figure 5a. Fast Fourier transform patterns were obtained from three regions on the HRTEM image, outside of the loop (blue), near the center of the loop (orange), and on the edge of the loop (red), shown in figure 5 b-d. FFT was taken at regions outside of the loop (blue) and on the edge of the loop (red) show a pure hexagonal GaN pattern, highlighted with red circles. FFT near the center of the loop shows additional patterns, highlighted with yellow circles that correspond to cubic GaN, where $[110]_{cubic}$ is parallel to $[११\underline{2}0]_{hexagonal}$. Iwata et al., have reported similar dislocation loops in Mg implanted GaN and showed these loops are vacancy type dislocation loops [14]. Our earlier study showed that during post-implantation annealing, the removal of implant-induced strain is accompanied by lattice rearrangement such as the generation of stacking faults, which can lead to the formation of cubic phase [15]. We speculate that a similar process occurs during the formation of these dislocation loops. Lattice rearrangement disrupts the original hexagonal stacking and results in parts of the lattice exhibiting cubic stacking. On the other hand, there is an overall decreasing trend in the defect density after annealing at a higher temperature or for a longer time. The loop defect density was measured to be $3.8 \times 10^9$ cm$^{-2}$ for 1400 °C 10 min case, $1.5 \times 10^9$ cm$^{-2}$ for the 1400 °C 100 min case, and $1.1 \times 10^9$ cm$^{-2}$ for 1500 °C 10 min case. Therefore, annealing at high temperature (≥ 1400 °C) showed no sign of Mg segregation and a decreasing trend in defect density which is expected to improve device performance.

**Conclusion**

The defect characteristics in magnesium implanted homoepitaxial GaN with ultra-high $N_2$ pressure annealing were investigated. Annealing of Mg implanted GaN with one GPa $N_2$ pressure at 1300 °C for 10 minutes completely removed the implant-induced strain. Extended defects remain after annealing with both pyramidal and trapezoidal inversion domains, as well as dislocation loops. The inversion domains show clear Mg segregation and produce electrically inactive Mg that limits the dopant activation efficiency. Increasing the annealing temperature ≥ 1400 °C under the same condition results in a decrease in residual loop defect density with no sign of Mg segregation at these defects. Pyramidal and trapezoidal inversion domains also do not exist with annealing temperatures ≥ 1400 °C. This study provides the structural analysis of the residual defects, showing Mg segregation in PIDs and TIDs is detrimental to dopant activation and can be avoided if the post-implant annealing temperature is ≥ 1400 °C. This work complements an earlier study with dopant activation efficiency data showing that annealing at 1400 °C leads to higher dopant activation compared to annealing at 1300 °C. Results from this work are expected to help achieve higher activation efficiency of p-type doping for devices including vertical GaN device structures.

**Acknowledgments**

This research was supported through the ARPA-E PNDIODES program under contract DE-AR0001116 at UCLA. This research used resources of the Advanced Photon Source, a U.S. Department of Energy (DOE) Office of Science User Facility operated for the DOE Office of Science by Argonne National Laboratory under contract no. DE-AC02-06CH11357. The synchrotron x-ray topography measurements were carried out at 1-BM beamline of the Advanced Photon Source, Argonne National Laboratory. TEM samples preparation and measurements were performed at the Center for Nanoscale Materials, Argonne National Laboratory.

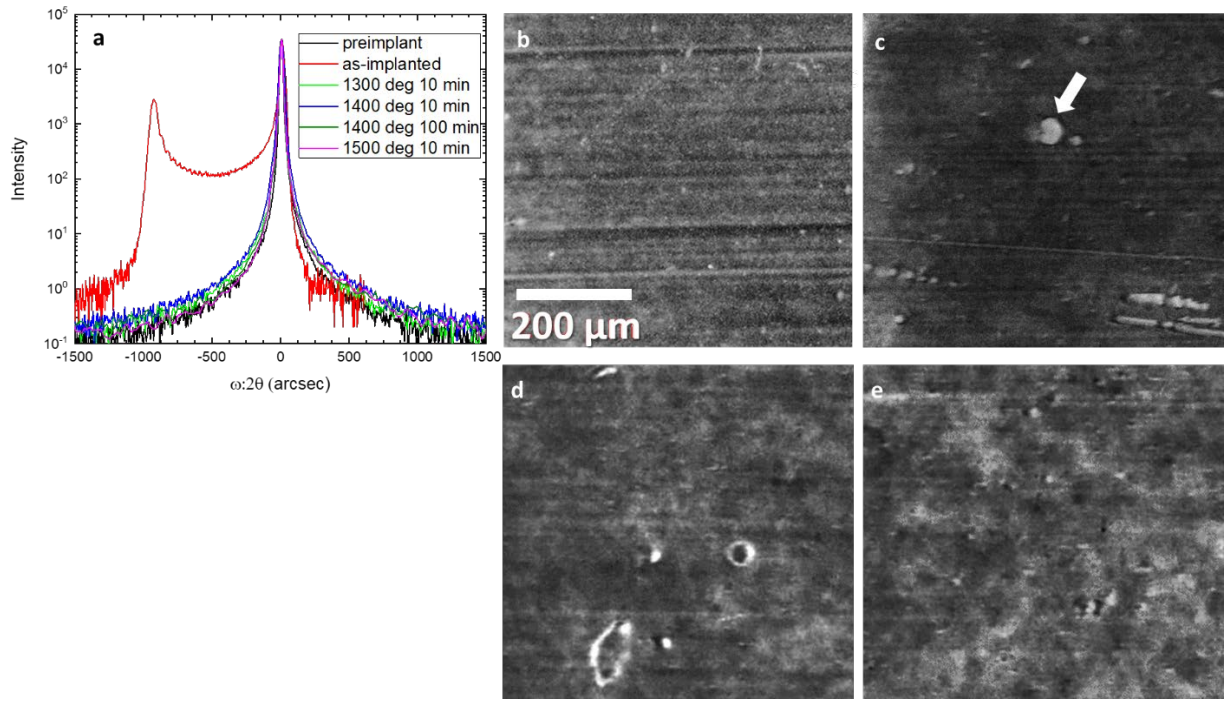

Figure 1. (a) Triple axis X-ray ω:2θ line scans near GaN (0004) peak for the samples showing the implant induced strain was fully relieved after annealing at 1300 °C for 10 minutes and annealing at temperatures ≥ 1400 °C has no further impact on the strain state; X-ray topography images exposed at a single point along a rocking curve for: (b) as-implanted; (c) annealed at 1300 °C for 10 minutes; (d) annealed at 1400 °C for 100 minutes, and (e) annealed at 1500 °C for 10 minutes. The white dots that appeared in all 4 images (b-e) are individual dislocations, and the densities are on the order of $10^4$ cm$^{-2}$.

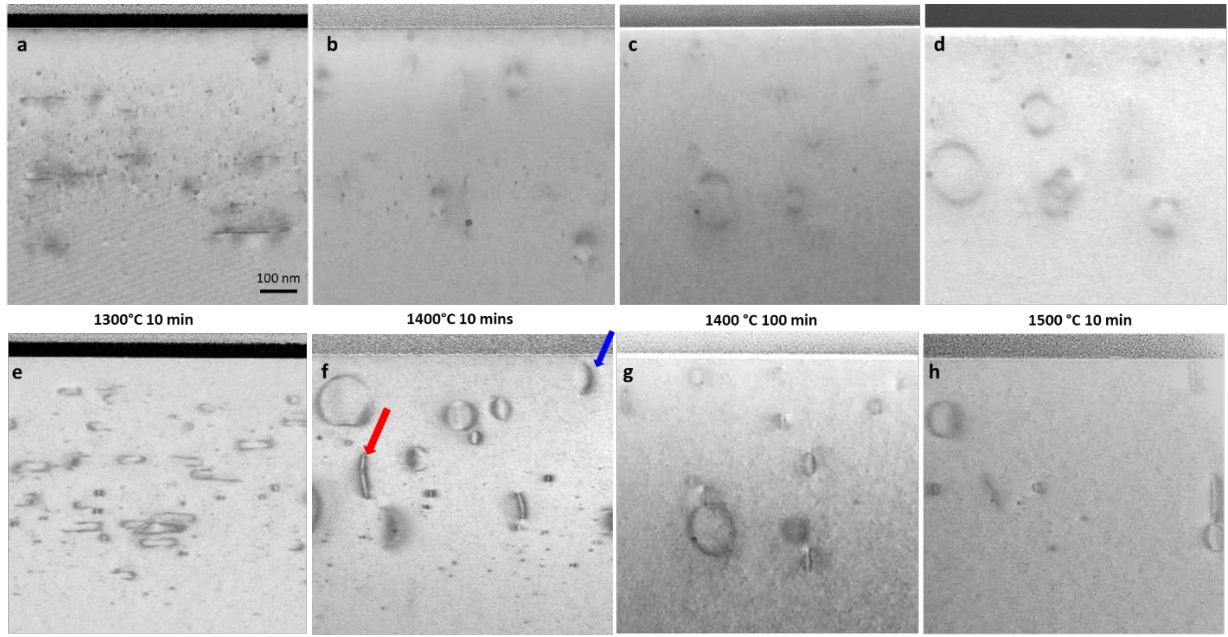

Figure 2. Cross-section STEM two beam condition bright field images with diffraction vector $g$ = <0002>, (a) 1300 °C annealed 10 min sample; (b) 1400 °C annealed 10 min sample; (c) 1400 °C annealed 100 min sample; (d) 1500 °C annealed 10 min sample; and diffraction vector $g$ = <11$\bar{2}$0>; (e) 1300 °C annealed 10 min sample; (f) 1400 °C annealed 10 min sample; (g) 1400 °C annealed 100 min sample; (h) 1500 °C annealed 10 min sample. In (f), the arrows show examples of dislocation loops that are not parallel to the FIB cut. All images share the same scale bar shown in (a).

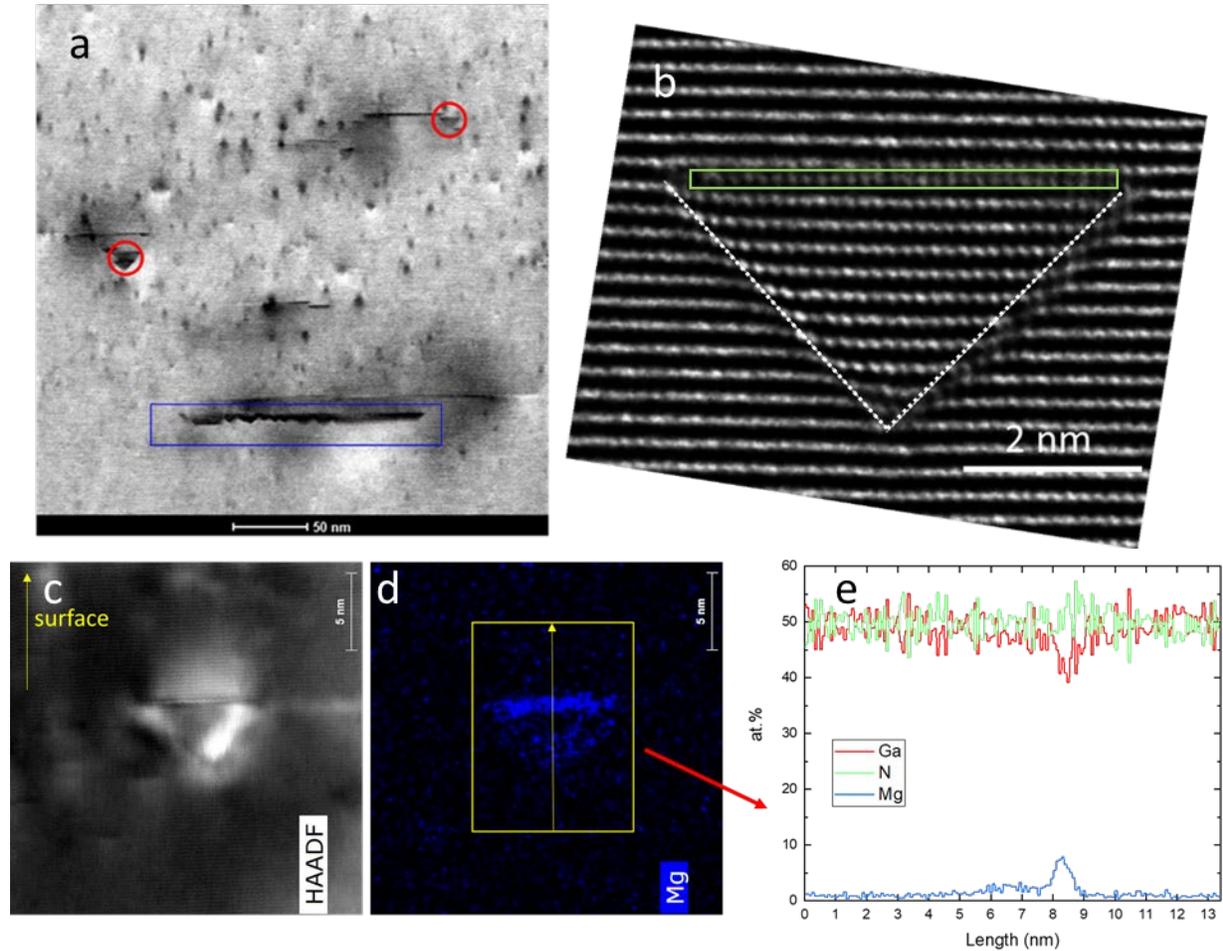

Figure 3. (a) STEM image of the 1300 °C 10 min sample showing examples of pyramidal inversion domain (circled in red) and trapezoidal inversion domain (boxed in blue); (b) HRTEM image of a pyramidal inversion domain showing the {11$\bar{2}$3} facets (highlighted with dotted white lines) and the extra layer of atoms near the base (boxed in green); (c) HAADF image of a pyramidal inversion domain; (d) EDX map showing the Mg segregation in the pyramidal inversion domain shown in c. (e) EDX line profile generated by integrating the intensity in the yellow box in Fig. 3d (the arrow shows the direction of the line profile, i.e. the start is at the bottom).

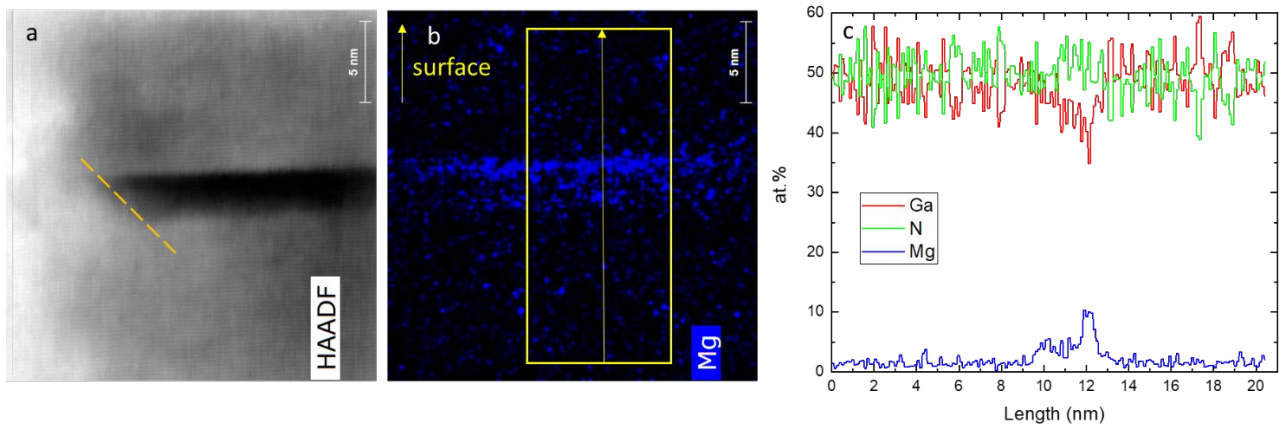

Figure 4. (a) HAADF image of a trapezoidal inversion domain with the edge also showing {112̱3} facets (highlighted with a diagonal orange line); (b) EDX map showing the Mg segregation in the trapezoidal inversion domain shown in a. (c) EDX line profile generated by integrating the intensity in the yellow box in Fig. 4b (the arrow shows the direction of the line profile, i.e. the start is at the bottom).

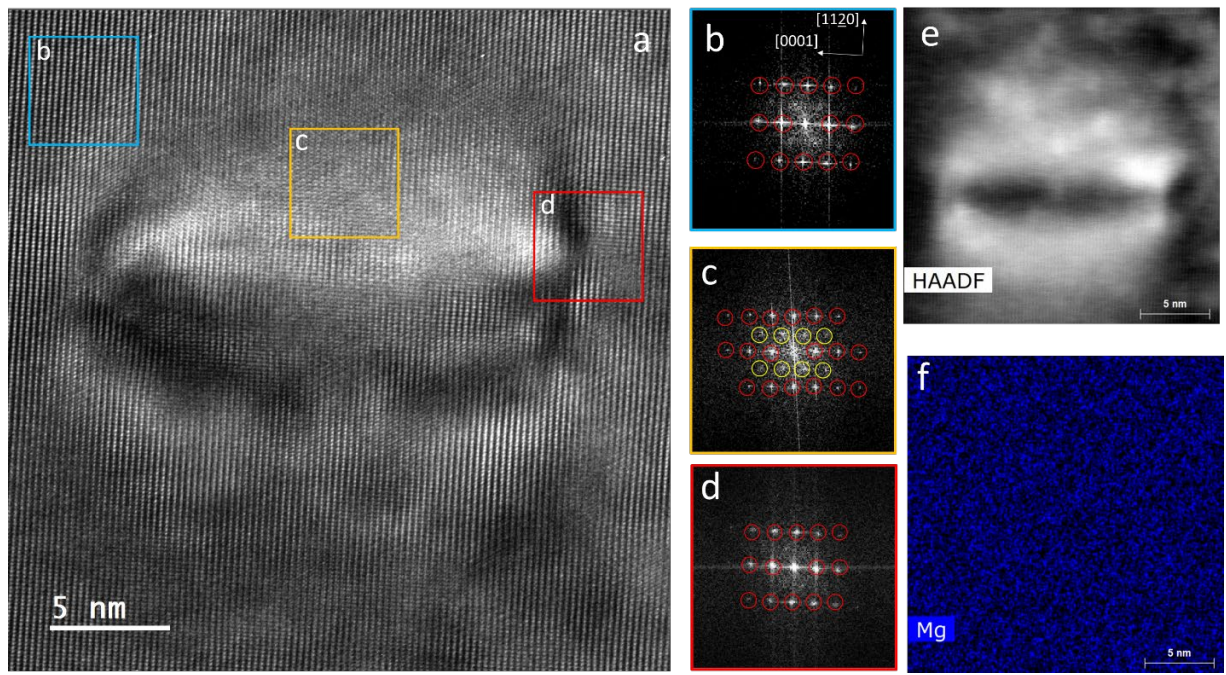

Figure 5. (a) HRTEM image of a small dislocation loop, fast Fourier transform pattern from three regions (three boxes) are shown in b-d; b) FFT outside of the loop showed pure hexagonal pattern (red circles); c) FFT close to the center of the loop show hexagonal GaN plus additional cubic GaN pattern (yellow circles); d) FFT on the edge of the loop also show a pure hexagonal pattern; (e) HAADF image of a small dislocation loop and the (f) the EDX map showing no signs of the Mg segregation in the loop.